\documentclass[journal,11pt,onecolumn,draftclsnofoot,]{IEEEtran}

\usepackage{tikz}
\usetikzlibrary{shapes,arrows}
\usepackage{caption}
\usepackage{amsmath}
\usepackage{float}
\usepackage{cleveref}
\usepackage{cases}
\usepackage{algorithm}
\usepackage{algpseudocode}

\begin{document}
	\title{Fluid Dynamics-Based Distance Estimation Algorithm for Macroscale Molecular Communication}
	
	\author{Fatih~Gulec, Baris~Atakan\thanks{This work was supported by the Scientific and Technological Research Council of Turkey (TUBITAK) under Grant 119E041.}
		\thanks{The authors are with the Department
			of Electrical and Electronics Engineering, Izmir Institute of Technology, 35430, Urla, Izmir, Turkey. (email: fatihgulec@iyte.edu.tr; barisatakan@iyte.edu.tr)}
		\thanks{\textbf{This paper is submitted to IEEE Transactions on Nanobioscience on March 13, 2020.}}}%
	
	\maketitle
	
	\begin{abstract}
		Many species, from single-cell bacteria to advanced animals, use molecular communication (MC) to share information with each other via chemical signals. Although MC is mostly studied in microscale, new practical applications emerge in macroscale. It is essential to derive an estimation method for channel parameters such as distance for practical macroscale MC systems which include a sprayer emitting molecules as a transmitter (TX) and a sensor as the receiver (RX). In this paper, a novel approach based on fluid dynamics is proposed for the derivation of the distance estimation in practical MC systems. According to this approach, transmitted molecules are considered as moving droplets in the MC channel. With this approach, the Fluid Dynamics-Based Distance Estimation (FDDE) algorithm which predicts the propagation distance of the transmitted droplets by updating the diameter of evaporating droplets at each time step is proposed. FDDE algorithm is validated by experimental data. The results reveal that the distance can be estimated by the fluid dynamics approach which introduces novel parameters such as the volume fraction of droplets in a mixture of air and liquid droplets and the beamwidth of the TX. Furthermore, the effect of the evaporation is shown with the numerical results.
	\end{abstract}
	
	
	\IEEEpeerreviewmaketitle
	
	\section{Introduction}
	\IEEEPARstart{I}{n} molecular communication (MC), which is inspired by nature, chemical signals are employed instead of electrical signals for information transfer. The motivation for the emergence of MC is to establish communication between nanomachines in a nanonetwork \cite{nakano2013molecular, Atakan-2014}. However, MC has a potential for practical applications in macroscale \cite{Farsad-2016}.
	
	The first experimental study about macroscale MC is given in \cite{farsad2013tabletop} where a MC link between a sprayer and a sensor can be established by using alcohol molecules. \cite{farsad2017novel} proposes an experimental MC system which is inspired by the human cardiovascular system for macroscale MC. This system uses  the pH level of a chemical  to encode information symbols. In \cite{unterweger2018experimental}, a similar system is proposed where magnetic nanoparticles are employed instead of chemicals. In \cite{giannoukos2017molecular} and \cite{mcguiness2018parameter}, it is shown that an odor generator and a mass spectrometer can be employed as a molecular transmitter (TX) and molecular receiver (RX), respectively. Multiple-input multiple-output (MIMO) technique is proposed in \cite{koo2016molecular} to increase the data rate for macroscale MC. \cite{zhai2018anti} proposes a method to mitigate inter-symbol interference for a practical macroscale MC system. Furthermore, a bio-inspired target detection algorithm is proposed in \cite{zhai2018bio} for the autonomous  movement of mobile RX robots towards a static TX in a mobile macroscale MC system.
	
	In MC, the distance between the TX and RX is a significant channel parameter, since higher data rates can be achieved via an accurate distance estimation  by arranging the MC system parameters properly \cite{atakan2007information, nakano2013transmission}. Furthermore, the location of a molecular TX in a molecular network can be found by an accurate distance estimation in a practical application. For instance, the location of an infected human which spreads a disease by emitting aerosol droplets through sneezing or coughing in a crowded place can be predicted by employing biological sensors as the RX and the infected human as the TX \cite{khalid2019communication}.  The distance estimation methods in the literature can be discussed as two-way and one-way methods. In two-way methods, the TX sends a molecular pulse signal to the RX and the RX sends a feedback signal to the TX, when it is received. Subsequently, the distance is calculated by the TX based on this feedback signal  \cite{moore2010measuring,moore2012measuring,moore2012comparing}. In one-way distance estimation methods which need less time with respect to two-way methods, the distance is estimated by the RX. In \cite{huang2013distance}, the received peak concentration or the peak time between the consecutive transmissions are measured by the RX for distance estimation in a 1-D channel. In \cite{wang2015distance}, two distance estimation schemes are proposed for a 3-D diffusion channel where the received peak time and the received energy are used for the first and second scheme, respectively. In addition, an algorithmic distance estimation method is proposed in \cite{wang2015algorithmic}  by the same authors. The study in \cite{lin2019high} proposes a method which uses maximum likelihood estimation for a 3-D diffusion channel. \cite{noel2015joint} proposes joint estimation methods for channel parameters  in a diffusive channel with a steady flow and degradable molecules. In all of these estimation methods, an ideal microscale channel where molecules propagate via diffusion is considered. Our study  given in \cite{gulec2019distance} proposes five methods including three data analysis based and two machine learning methods for distance estimation in a practical macroscale MC system for the first time. However, the physical meanings of the estimation parameters on which these methods depend are not known.
	
	%Furthermore, the location of a TX in a molecular network can be found by an accurate distance estimation in a practical application.
	
	In fact, the liquid is sprayed as droplets rather than molecules in practical macroscale scenarios \cite{ghosh1994induced,al2014influence}. Due to the initial velocity of droplets, they interact with air particles during their propagation. Therefore, considering only the diffusion of molecules is not sufficient, and a fluid dynamics perspective is needed for a more accurate distance estimation in macroscale scenarios as initially discussed in \cite{gulec2019distance}. In this paper, the Fluid Dynamics-Based Distance Estimation (FDDE) algorithm is proposed for a practical macroscale MC system. In the FDDE algorithm, droplets are considered as information carriers in the channel and a two-phase flow model is used. Here, liquid phase of droplets and gas phase of air particles represent these two phases. In this model, liquid droplets and gaseous air particles move together as a mixture. Moreover, it is considered that droplets evaporate as the time elapses. The TX is modeled as a directed emitter with a predefined beamwidth. Droplets are assumed to move in a cone shaped volume determined by this beamwidth. Then, the laws of mass and momentum conservation are utilized to estimate the average propagation distance of the droplets from the TX by analytical derivations. The FDDE algorithm employs these derivations by also considering reducing droplet diameter due to  evaporation as the droplets propagate in the MC channel. Subsequently, the proposed FDDE algorithm is validated by experimental data. It is shown that the distance between the TX and RX in the MC channel depend on novel parameters such as the beamwidth of the TX, volume fraction of the droplets in the mixture volume and densities of the air particles and liquid droplets. Moreover, the results show that the effect of the evaporation for shorter distances is small. With the validation of the FDDE algorithm, it is revealed that modeling the motion of the droplets with a fluid dynamics approach can be employed  for  distance estimation in practical macroscale MC systems. 
	
	The rest of the paper is organized as follows. The experimental setup is given in Section \ref{Exp_Setup}. The proposed FDDE algorithm is presented in Section \ref{Estimation}. The experimental work explaining the measurement of the required parameters is given in Section \ref{Measurements}. The numerical results including validation of the proposed FDDE algorithm are presented in Section \ref{Numerical_3}. Finally, concluding remarks are given in Section \ref{Conclusion}.
	\section{Experimental Setup}
	\label{Exp_Setup} 
\begin{figure}[h]
		\centering
		\scalebox{0.195}{\includegraphics{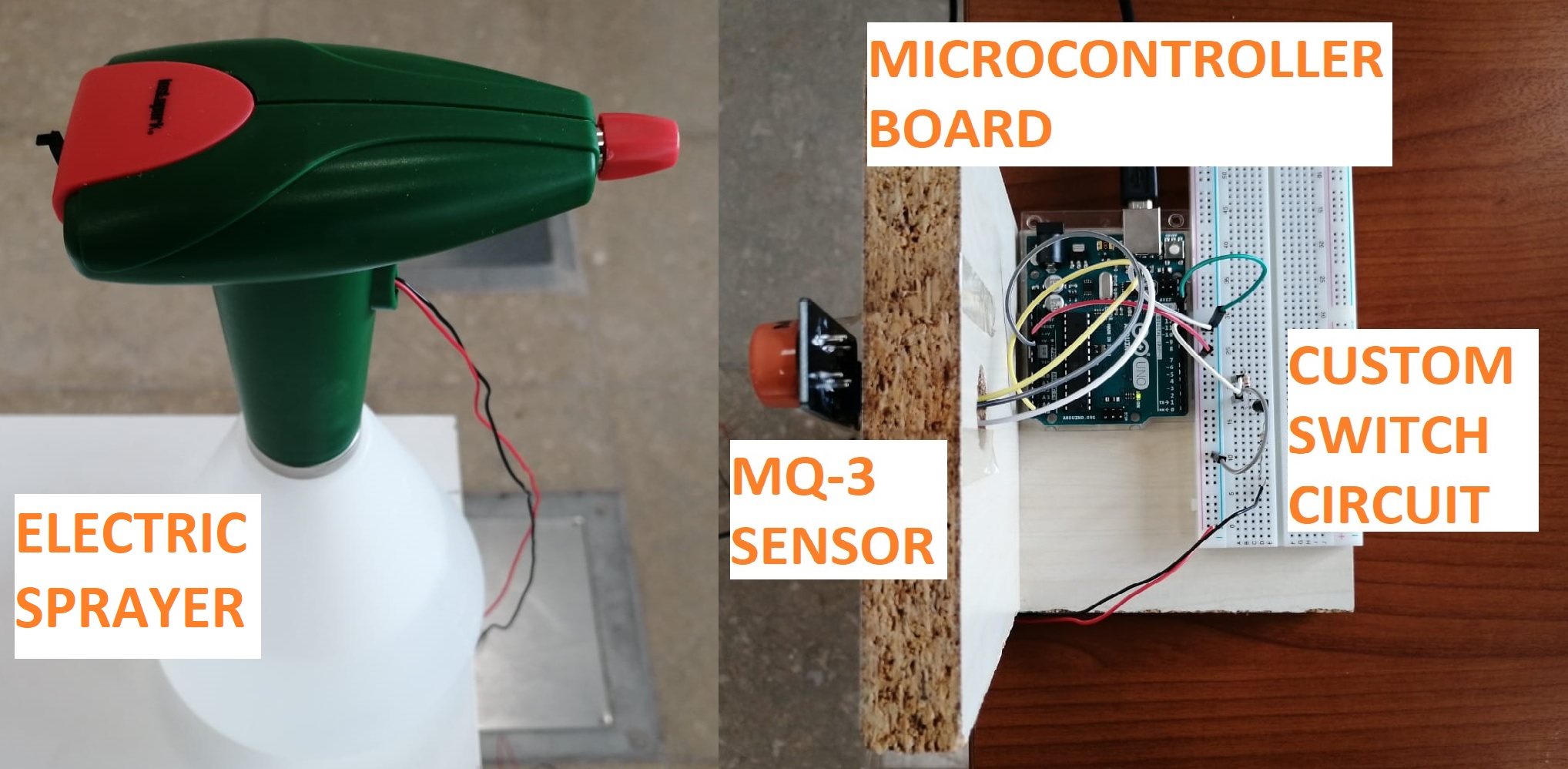}}
		\caption{The experimental setup: The TX (left) and RX (right).}
		\label{Components}
	\end{figure}

	The experimental setup given in Fig. \ref{Components} includes a TX, a RX and the propagation channel (air). The TX is an electric sprayer which transmits ethanol droplets with an initial velocity and the RX is an MQ-3 alcohol sensor. In order to establish synchronization, the TX and RX are controlled by the same Arduino microcontroller which is connected to a computer. The TX is connected to the microcontroller via a custom switch circuit. In this study, the TX emits the ethanol droplets for different distances between the TX and RX as a short pulse whose duration is defined as the emission time ($T_e$). No fan is used to generate a flow between the TX and RX. Furthermore, the alignment of the TX and RX on the horizontal axis is made precisely. The experimental setup detailed in this section is employed for the validation of the proposed FDDE algorithm, which is explained in the next section.
	\section{Fluid Dynamics-Based Distance Estimation Algorithm}\label{Estimation} 
\begin{figure}[t!]
		\centering
		\scalebox{0.7}{\includegraphics{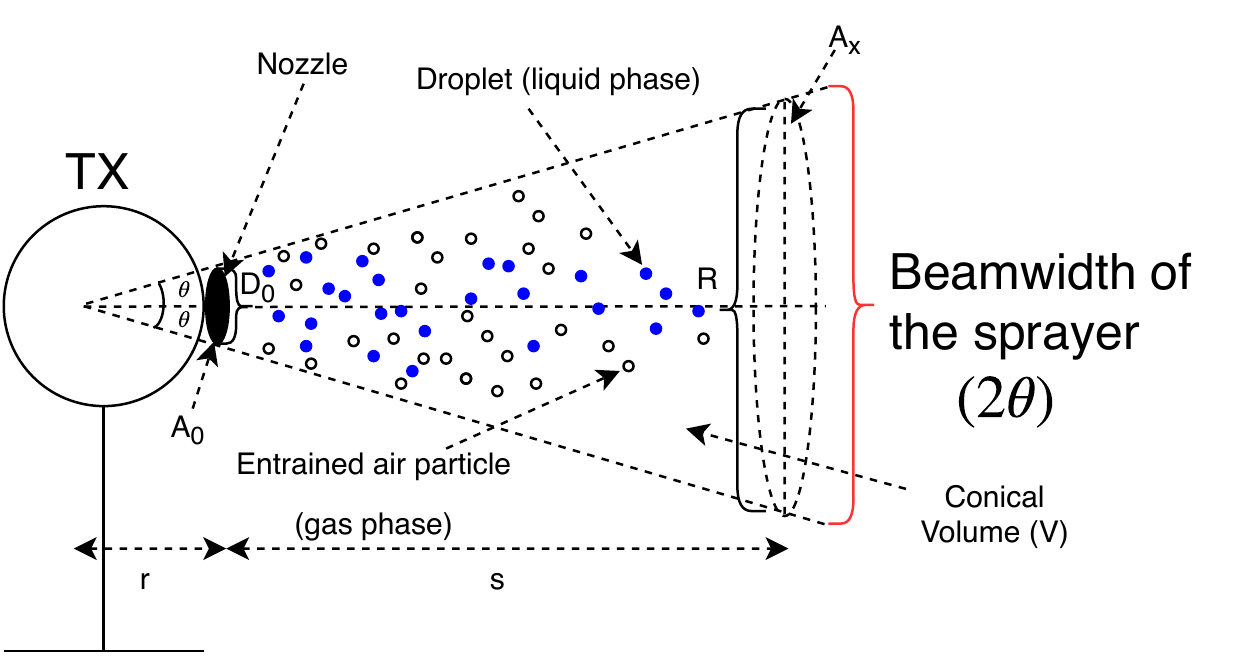}}
		\caption{Two-Phase Flow Model.}
		\label{Two-phase}
	\end{figure}

	In this section, the FDDE algorithm is proposed to estimate the distance between the TX and RX with a fluid dynamics approach. When the droplets are transmitted into the air, air particles are entrained by the droplets and this induces an air flow. During this interaction generating this air flow in the vicinity of the nozzle of the TX, the velocity difference among the droplets and air particles is large and this difference fluctuates over time. Here, it is useful to introduce Reynolds number ($Re$) which determines the flow type of the fluid, i.e., laminar or turbulent. The changing relative velocity of the droplets with respect to air particles cause $Re$ to grow which is given by
	\begin{equation}
	Re = \frac{d |v_d - v_a|}{\nu_a}
	\label{Re}
	\end{equation}
	where $d$ is the diameter of the droplet, $\nu_a$ is the kinematic viscosity of the air, $v_d$ and $v_a$ are the velocities of the droplets and air particles, respectively. As $Re$ increases, deriving analytical solutions for the motion of particles becomes difficult, since the turbulent diffusivity increases \cite{ghosh1994induced}. However, as the distance between the TX and RX increases, the droplets can be assumed to have nearly the same velocity with the entrained air particles. This situation makes $Re$ small and turbulent flows are not considered. Hence, a tractable analytical solution for the average velocity and traveling distance of the droplets is feasible. For a macroscale scenario without a constant flow, the distance between the TX and RX is sufficiently large that the effect of the initial interaction among the droplets and air particles can be negligible for the total traveling distance of the droplets. The motions of the droplets and air particles with the same velocities can be modeled by using the two-phase flow model. In this model, the two phases represent the liquid and gas phases of the fluid particles, i.e., liquid droplets and gaseous air particles. As illustrated in Fig. \ref{Two-phase}, droplets and air particles move together as a mixture between the TX and RX. This mixture is assumed to propagate in a cone shaped volume which has a beamwidth of $2\theta$ (see Fig. 2). In fluid dynamics literature, two-phase flow models are applied to estimate the average distance of the fuel droplets sprayed by a fuel injector in a car engine \cite{sazhin2001model}  or the coverage of the sprayed pesticide droplets in agriculture \cite{al2014influence}. In this study, the two-phase flow model given in \cite{sazhin2001model} is adopted and modified to estimate the propagation distance of the droplets transmitted from the TX. Furthermore, the evaporation of the droplets, which is detailed as follows, is considered for the FDDE algorithm.
	\subsection{Evaporation of Droplets}
	The transmitted droplets can evaporate as they move away from the sprayer \cite{al2014influence}. The evaporation can be described in terms of the reduction in the droplet diameter. The diameter change with respect to time is modeled as \cite{mokeba1997simulating}
	\begin{equation}
	\frac{\partial d}{\partial t} = -2\frac{M_v D_v \rho_a \Delta P }{M_a d \rho_d P_a} ( 2 +0.6 Sc^{1/3} Re^{1/3}),
	\label{delta_d}
	\end{equation}
	where $M_v$ is the molecular weight of the evaporating vapor, $M_a$ is the molecular weight of the air, $D_v$ is the diffusion coefficient of the droplet's vapor in the air, $d$ is the droplet diameter, $P_a$ is the partial air pressure, $\Delta P$ is the pressure difference between the droplet surface and the diffusing vapor in the air and $Sc$ is the Schmidt number which is given by $Sc = \nu_a/D_v$. The parameters and subscripts used for the proposed algorithm are summarized in Table \ref{Definitions}.
	\begin{table}[h]
		%	\vspace{0.5cm}
		\centering
		\caption{Definitions of the parameters and subscripts}
		\scalebox{1}{
			\begin{tabular}{p{40pt}|p{120pt}|p{40pt}|p{120pt}}
				\hline
				\textbf{Parameter}	& \textbf{Definition} & \textbf{Parameter}	& \textbf{Definition}\\
				\hline  \hline 
				$Re$ & Reynolds number & $M$ & Molecular weight \\
				$\theta$  & Half-beamwidth of the TX &  $d$ & Average diameter of the droplet\\
				$D$ & Diffusion coefficient & $\rho$ & Density \\
				$\Delta t$  &  Time step  & $P$ & Partial pressure \\
				$Sc$ & Schmidt number & $m$ & Mass\\
				$\Delta P$ & Vapor pressure difference & $V$ & Conical mixture volume\\ 
				$\alpha_{d}$ & Volume fraction of droplets in mixture & $T_e$ &  Emission time of the TX \\
				$D_0$  & Diameter of the nozzle orifice & $A$ & Circular cross-sectional area \\ 
				$v$ & Average velocity & $ \dot{m} $ & Mass flow rate\\			
				$s$ & Average distance & $R$ & Diameter of $A$ \\
				$p$ & Momentum & $t_s$ & Total simulation time\\
				$r$ & Distance between the nozzle and starting point of the flow & $\dot{p}$ & Time rate of change of the linear momentum\\					
				\hline \textbf{Subscript} & & \textbf{Subscript}	 \\ \hline 
				$d$ & droplet (liquid ethanol) & $i$ & value at the $i^{th}$ time step\\
				$v$ & vapor ethanol & $x$ & mixture\\
				$a$ & air &	$0$ & initial value\\
				\hline   \hline           
		\end{tabular}}
		\label{Definitions}
	\end{table}
	
	For an algorithm which evaluates the motion of droplets by using (\ref{delta_d}), the diameter change ($\Delta d$)  in every time step can be given as below
	\begin{align}
	d_{i} &= d_{i-1} + \Delta d \\
	\Delta d &= -2\frac{M_v D_v \rho_a \Delta P }{M_a d_i \rho_d P_a} ( 2 +0.6 S_c^{1/3} Re^{1/3}) \Delta t,
	\label{d_i}
	\end{align} %
	where  $d_i$ is the diameter at the time $t_i$ which increases with steps of $\Delta t$ for each iteration and represents the elapsed time at the $i^{th}$ time step. In order to estimate the distance values in a practical MC system, it is essential to relate these values with the  measured sensor voltage values. For this purpose, $t_i$ is assumed as the peak time of the signal measured by the sensor, since the majority of droplets are assumed to reach the RX at this peak time.
	
	Here, a new parameter ($\alpha_d$) which shows the volume fraction of the droplets in the mixture of droplets and air particles (see Fig. \ref{Two-phase})  is introduced. The evaporation of droplets causes $\alpha_d$ to be time-dependent. As the diameters of droplets change with time, $\alpha_d$ is needed to be updated. To derive a model for this, let $N$ be the number of  droplets, $d_i$ the average diameter of a spherical droplet, $V$  the mixture volume and $\alpha_{d_i}$  the volume fraction of droplets in the mixture volume during the $i^{th}$ time step. Here, $V$ is considered as a constant cone-shaped volume with the beamwidth of the TX for the distance between the TX and RX as shown in Fig. \ref{Two-phase}. Then, the volume fractions for consecutive time steps can be given as
	\begin{align}
	\alpha_{d_{i-1}} &= \dfrac{N \frac{4}{3} \pi \left(\frac{d_{i-1}}{2}\right)^3}{V} \\
	\alpha_{d_i} &= \dfrac{N \frac{4}{3} \pi \left(\frac{d_i}{2}\right)^3}{V}.
	\end{align}
	%Since the traveled distance is very small for a very small time step, it can be assumed that $V_i \approx V_{i+1}$. 
	Hence, the relation between consecutive volume fractions of droplets is given by
	\begin{equation}
	\alpha_{d_i} = \alpha_{d_{i-1}} \left(\frac{d_i}{d_{i-1}}\right)^3.
	\label{alpha_d1}
	\end{equation}
	For this derivation, it is essential to consider that the number of droplets in the mixture volume increases during the emission of droplets. Therefore, it is assumed that the volume fraction of droplets at $t = T_e$ is a pre-known constant, i.e., $\alpha_{d_0}$, and  $ \alpha_{d_i} $ increases linearly between $t = 0$ and $ t= T_e$. After $t = T_e$, the volume fraction is given by (\ref{alpha_d1}). Hence, the volume fraction of droplets can be defined as a time-dependent function:
	\begin{numcases}
		{\alpha_d(t = t_i) = \alpha_{d_i} =} 
		\frac{\alpha_{d_0}}{T_e} t_i &\hspace{-5mm}$, 0\leq t_i\leq T_e $ \label{alpha_d_i0}\\ 
		\alpha_{d_{i-1}} \left(\frac{d_{i}}{d_{i-1}}\right)^3 &\hspace{-5mm}$, t_i > T_e $. \label{alpha_d_i}
	\end{numcases}
	\subsection{Propagation of Evaporating Droplets in Two-Phase Flow} 
	In order to  clearly explain the propagation of evaporating droplets in  two-phase flow, some definitions related with fluid dynamics are  first given as follows. As illustrated in Fig. \ref{Two-phase}, the liquid phase and the gas phase form two phases of the two-phase flow model in which liquid droplets and gaseous air particles move together as a mixture. This model can be generated by employing the laws of mass and momentum conservation. In fluid dynamics, the conservation of mass is applied by using the mass flow rate ($ \dot{m} $) which is defined as the amount of mass flowing through a surface per unit time (kg/s) \cite{munson2009fundamentals}. From the conservation of mass, which states that the net mass flow rate ($ \dot{m} $) is zero in a closed system (a system with no external forces acted on and no external matter exchange), the equation below can be written \cite{munson2009fundamentals}
	\begin{equation}
	\dot{m_0} = \dot{m_i},
	\label{m_12} 
	\end{equation}
	where $ \dot{m_0} $ and $ \dot{m_i} $  is the mass flow rate of droplets for the initial state and  $i^{th}$ time step, respectively. The mass flow rate is expressed by \cite{munson2009fundamentals}
	\begin{equation}
	\dot{m}_i = \rho_i A_i v_i,
	\label{dotm}
	\end{equation}
	where $\rho_i$ is the density of the fluid (kg/m$^3$) and $v_i$ is the fluid velocity (m/s) which is perpendicular to the cross-sectional area of the fluid $A_i$ (m$^2$) for the $i^{th}$ time step. The mass flow rate of droplets at the $i^{th}$ time step can be expressed by subtracting the mass flow rate of air particles from the mixture of droplets and air particles. Then, (\ref{m_12}) can be given as 
	\begin{equation}
	\rho_d A_0 v_0 = \rho_{x_i} A_{x_i} v_{x_i} - (1-\alpha_{d_i}) \rho_a A_{x_i} v_{x_i},
	\label{Mass}
	\end{equation}
	where $A_0$ is the circular cross-sectional area of the nozzle, $v_0$ is the initial velocity of droplets, $ \rho_{x_i} $ is the density of the mixture, $v_{x_i}$ is the velocity of the mixture and $A_{x_i}$ is the circular cross-sectional area of the mixture volume at the $i^{th}$ time step as illustrated in Fig. \ref{Two-phase}. In (\ref{Mass}), the term in the left-hand side shows the initial mass flow rate of droplets. The first term and second term of the right-hand side  gives the mass flow rate of the mixture and entrained air, respectively.
	
	As observed in Fig. \ref{Two-phase}, the geometrical relation between $A_{x_i}$ and $A_0$ can be represented as given by
	\begin{equation}
	\tan(\theta) = \frac{D_0}{2r} = \frac{R_i}{2(s_i+r)},
	\label{cone}
	\end{equation}
	where $D_0$ is the diameter of the nozzle orifice, $r$ is the distance between the nozzle and the assumed theoretical starting point of the flow, $\theta$ is the half-beamwidth of the TX, $s_i$ is the distance traveled by the mixture or droplets and $R_i$ is the diameter of $A_{x_i}$ at the $i^{th}$ time step. By using (\ref{cone}), $A_{x_i}$ can be represented in terms of $A_0$, $s_i$, $D_0$ and $\theta$ as given by
	\begin{equation}
	A_{x_i} = A_0 + \pi D_0\hspace{0.5mm} s_i\hspace{0.5mm} \tan(\theta) + \pi s_i^2 \tan^2(\theta).
	\label{A_x}
	\end{equation} 
	
	Next, the conservation of the momentum is exploited to derive the estimated distance. The momentum ($p$) can be defined for the $i^{th}$ time step as given by \cite{munson2009fundamentals}
	\begin{equation}
	p_i = m_i v_i.
	\label{mom}
	\end{equation}
	The law of momentum conservation states that the total momentum does not change for a closed system. In our model, it is assumed that the fluid motion is linear (not rotational) and only on the horizontal axis. As in the case of conservation of mass, the time rate of change of the linear momentum, which is defined as the change of the momentum in time, is employed to apply the momentum conservation in fluid dynamics \cite{munson2009fundamentals}. To that end, the linear momentum equation, which implies that the system's time rate of change of the linear momentum ($\dot{p}$) is constant for a closed system, is used  as given by \cite{munson2009fundamentals}
	\begin{equation}
	\dot{p}_0 = \dot{p}_i,
	\label{dotp}
	\end{equation}
	where the left-hand side and right-hand side statements are the time rate of changes of the linear momentum at the initial state of droplets and the mixture state, respectively. By substituting (\ref{mom}) into (\ref{dotp}), the linear momentum equation is expressed in terms of mass flow rate as given by \cite{munson2009fundamentals}
	\begin{align}
	\dot{m_0} v_0 &= \dot{m_i} v_i \label{dotp2}\\
	\rho_d A_0 v_0^2 &= \rho_{x_i} A_{x_i} v_{x_i}^2,
	\label{Momentum}
	\end{align}
	where the mass flow rate derived in (\ref{dotm}) is substituted into (\ref{dotp2}). The equations (\ref{Mass}), (\ref{A_x}) and (\ref{Momentum}) are manipulated to derive the traveling distance of the mixture. Firstly, (\ref{Mass}) is multiplied by $1/(A_{x_i} v_{x_i})$ and $\rho_{x_i}$ is derived as shown below.
	\begin{equation}
	\rho_{x_i} = \frac{\rho_d A_0 v_0}{A_{x_i} v_{x_i}} + (1-\alpha_{d_i}) \rho_a.
	\label{abbr1}
	\end{equation}
	Secondly, (\ref{Momentum}) can be written as
	\begin{equation}
	\rho_{x_i} = \frac{\rho_d A_0 v_0^2}{A_{x_i} v_{x_i}^2}.
	\label{abbr2}
	\end{equation}
	Then, (\ref{abbr2}) is substituted into (\ref{abbr1}) and all the terms are taken to the left-hand side as given by
	\begin{equation}
	\frac{\rho_d A_0 v_0^2}{A_{x_i} v_{x_i}^2} - \frac{\rho_d A_0 v_0}{A_{x_i} v_{x_i}} - (1-\alpha_{d_i}) \rho_a = 0.
	\label{eq4}
	\end{equation}
	Afterwards, we let $\tilde{v}_i = v_0/v_{x_i}$ and write (\ref{eq4})  in terms of $\tilde{v}_i$ as
	\begin{equation}
	\tilde{v}_i^2 - \tilde{v}_i - \frac{(1-\alpha_{d_i}) \rho_a A_{x_i}}{\rho_d A_0} = 0.
	\label{eq4.2}
	\end{equation}
	For physically meaningful parameter values, one of the roots of (\ref{eq4.2}) is negative and the positive root is given by
	\begin{equation}
	\tilde{v}_i = \frac{v_0}{v_{x_i}}  = \frac{1}{2}\left( 1 + \sqrt{\frac{4 (1-\alpha_{d_i}) \rho_a A_{x_i}}{\rho_d A_0}} \right).
	\label{eq4.3}
	\end{equation}
	By using (\ref{eq4.3}), $v_{x_i}$ is found as
	\begin{equation}
	v_{x_i}  = \frac{2 v_0}{ 1 + \sqrt{\frac{4 (1-\alpha_{d_i}) \rho_a A_{x_i}}{\rho_d A_0}}}.
	\label{eq4.4}
	\end{equation}
	By substituting $A_0 = \frac{\pi D_0^2}{4}$ and (\ref{A_x}) into (\ref{eq4.4}), $v_{x_i}$ is derived as
	\begin{equation}
	v_{x_i}  = \frac{2 v_0}{ 1 + \sqrt{k_1 + k_2 s_i + k_3 s_i^2}}.
	\label{eq5}
	\end{equation}
	where
	\begin{multline}
	k_1 = 1 + 4 (1-\alpha_{d_i}) \frac{\rho_a}{\rho_d}, \hspace{0.2cm} k_2 = \frac{16 (1-\alpha_{d_i}) \rho_a \tan(\theta) }{D_0 \rho_d}, \\
	k_3 = \frac{16 (1-\alpha_{d_i}) \rho_a \tan^2(\theta)}{D_0^2 \rho_d}.
	\label{eq5.2}
	\end{multline}
	
	The average velocity of droplets can be expressed as the derivative of the distance with respect to time. Hence, $v_{x_i}$ is given in terms of the difference of the consecutive distance and time values as given by
	\begin{equation}
	v_{x_i} = \frac{\Delta s_i}{\Delta t} = \frac{s_i - s_{i-1}}{\Delta t},
	\label{eq6}
	\end{equation}
	where $\Delta s_i$ represents the average distance traveled by the droplets for the time duration $ \Delta t = t_i - t_{i-1} $. By incorporating (\ref{eq5.2}) into (\ref{eq5}), the following equation is derived: 
	\begin{equation}
	\sqrt{k_1 + k_2 s_i + k_3 s_i^2} = \frac{2 v_0 \Delta t}{\Delta s_i} - 1.
	\label{eq7}
	\end{equation}
	Then, both sides of (\ref{eq7}) are squared as given by
	\begin{equation}
	k_1 + k_2 s_i + k_3 s_i^2 = \frac{4 v_0^2 (\Delta t)^2 - 4v_0 \Delta t \Delta s_i + (\Delta s_i)^2}{(s_i - s_{i-1})^2}.
	\label{eq7.2}
	\end{equation}
	Next, when $ \Delta s_i = s_i - s_{i-1}$ is substituted into (\ref{eq7.2}), a quartic equation can be obtained as shown by
	\begin{multline}
	k_3 s_i^4 + (k_2 -2k_3 s_{i-1}) s_i^3 + (k_1 - 2k_2 s_{i-1} + k_3 s_{i-1}^2 - 1) s_i^2 \\ + (k_2 s_{i-1}^2 - 2k_1 s_{i-1}  + 4 v_0 \Delta t + 2 s_{i-1}) s_i + k_1 s_{i-1}^2 - 4v_0^2 (\Delta t)^2 \\ - 4 v_0 \Delta t s_{i-1} - s_{i-1}^2 = 0.
	\label{eq7.3}
	\end{multline}
	$s_i$ is found by solving the quartic equation in (\ref{eq7.3}) in terms of its previous value $s_{i-1}$ and $\Delta t$. For physically meaningful parameter values, one of the roots is negative and the two of the roots are complex numbers in (\ref{eq7.3}). Only one real positive root is left as the solution of $s_i$ which is too long to write in this paper. The results for this solution are shown numerically in the next section. 

	\begin{algorithm}
		\caption{FDDE Algorithm}
		\label{Alg1} 
		\begin{algorithmic}[1]
			\State \textbf{input:} $ T_e $, $ \alpha_{d_0}$, $D_0$ $v_0$, $\Delta t$, $\theta$, $t_s$
			\State $t_0 = 0$; $i = 1$; $d_0 = D_0$
			\State Calculate $\Delta d$ by (\ref{d_i})
			\While{$t_{i-1} \leq t_s$}
			\If{$t_{i-1} \leq T_e$}
			\State $d_i = d_{i-1}$
			\State Calculate $\alpha_{d_i}$ by (\ref{alpha_d_i0})
			\Else
			\State $d_i = d_{i-1} + \Delta d$ 
			\State Calculate $\alpha_{d_i}$ by (\ref{alpha_d_i})
			\EndIf
			\State Calculate $s_i$ by using the real positive root of (\ref{eq7.3})
			\State $ i = i+1 $; $ t_i = t_{i-1} + \Delta t$		
			\EndWhile
		\end{algorithmic} 
	\end{algorithm}
	
	Finally, an estimation algorithm given in Algorithm \ref{Alg1} is proposed by using the derived parameters to evaluate the average traveling distance of droplets. In this algorithm, the initial parameters are given at the beginning such as $T_e$, $\alpha_{d_0}$, $D_0$, $v_0$, $\Delta t$, $\theta$ and total simulation time ($t_s$). $D_0$ is assigned as the initial droplet diameter. Then, the diameter of droplets are calculated for each time step. During the emission of droplets, droplet diameters are assumed to be constant. Otherwise, the reduction in the droplet diameter is calculated by using $\Delta d$ to obtain $ \alpha_{d_i} $ as given in (\ref{alpha_d_i0}) or (\ref{alpha_d_i}). Subsequently, the average distance is calculated by using the real positive root of (\ref{eq7.3}). These steps are repeated until the end of the simulation. The FDDE algorithm can be utilized to model the movement of droplets emitted from the TX as given with the numerical results in the next section.
	\section{Numerical Results}\label{Exp_3}
	In this section, numerical results using the FDDE algorithm given in Section \ref{Estimation} are presented. This algorithm is validated by experimental results. Firstly, measurements of some practical parameters to be used as input in this algorithm are given.
	\subsection{Measurements}\label{Measurements}
\begin{figure}[H]
		\centering
		\scalebox{0.35}{\includegraphics{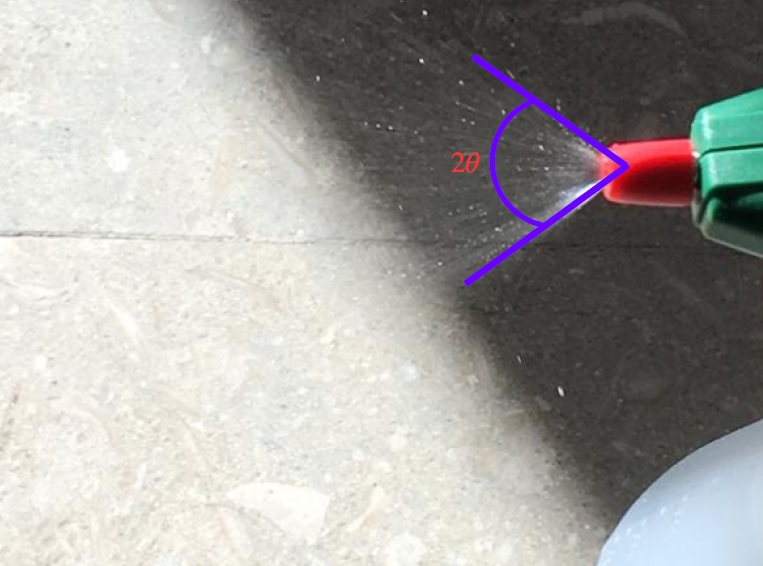}}
		\caption{Measurement for the beamwidth of the TX. }
		\label{Theta} 
	\end{figure}

	The parameters of the sprayer such as $\theta$, $D_0$ and $v_0$ are required to be measured to compare the practical experiment results with the proposed algorithm. $D_0$ is measured by a precise digital caliper, which is a measurement tool with a resolution of $0.01$ mm for the length of an object. $\theta$ is measured  by using image analysis with the software ImageJ. The image used for this analysis is given in Fig. \ref{Theta}.
	Then, the volumetric flow rate ($ Q $), which gives the fluid volume flowing through the sprayer per unit time, is measured in order to calculate the initial average velocity of droplets ($v_0$). For the measurement of $Q$, a precision balance, which is an equipment measuring the mass with a precision of $0.001$ g, is used. The mass of the sprayer filled with the liquid ethanol is measured before and after a short spraying. Meanwhile, the elapsed time for spraying is recorded. In order to find the volume of the sprayed liquid ethanol, the mass is divided by the density of the liquid ethanol ($\rho_d$) whose value is obtained from \cite{pubchem_ethanol}. Hence, $Q$ can be calculated  by dividing the mass difference to the elapsed time for consecutive measurements  \cite{munson2009fundamentals}. $Q$ can be presented by the formula
	\begin{equation}
	Q = \frac{\Delta V}{\Delta t_v},
	\label{Q}
	\end{equation}
	where $\Delta V$ and $\Delta t_v$ show the volume and time difference between the initial and final measurement values. In order to eliminate the random effects, ten measurements are made to calculate $Q$. Afterwards, the initial average velocity of the droplets can be found by \cite{munson2009fundamentals}
	\begin{equation}
	v_0 = \frac{Q}{A_0},
	\label{v_0}
	\end{equation}
	where $A_0$ is the cross-sectional area of the nozzle calculated by $A_0 = \pi (D_0/2)^2$. Thus, ten $v_0$ values are obtained as the result of these measurements. The average of these ten $v_0$ values is employed for the comparison of the theoretical and experimental results. The measured and calculated values of $\theta$, $D_0$ and $v_0$ are given in Table \ref{Sim_parameters}. 
	
	For the validation of the FDDE algorithm, the experimental data are collected by using the experimental setup given in Section \ref{Exp_Setup}. The measurements are made for nine different distances ranging from $ 1 $ to $ 1.8 $ m in steps of $ 0.1 $ m. At each distance, five different emissions are made with $T_e = 0.25$ s from the TX (a single puff) at $20^\circ$C. A sufficient amount of duration (at least five minutes) is left between two consecutive measurements in order to eliminate the effects of the previous transmissions. The room that the measurements are made is ventilated along this duration. In the next part, the experimental parameters given in Table \ref{Sim_parameters} are used for the numerical evaluation of the proposed methods.
	\begin{table}[H]
		%	\vspace{0.5cm}
		\centering
		\caption{Experimental parameters}
		\scalebox{0.85}{
			\begin{tabular}{ll}
				\hline
				\textbf{Parameter}	& \textbf{Value} \\
				\hline  \hline 
				Half-beamwidth of the TX ($\theta$)  & $ 38^\circ $ \\  
				Diameter of the nozzle orifice ($D_0$)  & $ 510$ $ \mu $m \\ 
				Average initial velocity of the droplets ($v_0$) & $ 10.8$ m/s \\
				Time step ($\Delta t$)  &  $ 0.01$ s  \\ 
				Reynolds number ($Re$) & $ 0.1 $ \\
				Emission time of the TX ($T_e$) &  $ 0.25$ s \\
				Volume fraction of droplets at $t=T_e$ ($\alpha_{d_0}$) & $ 0.001 $ \\
				Density of liquid ethanol ($\rho_d$) & $ 789$ kg/m$^3 $ \cite{pubchem_ethanol}\\
				Density of air  ($\rho_a$) & $ 1.2$ kg/m$^3 $ \cite{mokeba1997simulating}\\
				Molecular weight of ethanol ($M_v$) & $ 46.069 \times 10^{-3}$ kg/mol \cite{mokeba1997simulating}\\
				Molecular weight of air ($M_a$) & $ 28.9647 \times 10^{-3}$ kg/mol  \cite{pubchem_ethanol}\\
				Diffusion coefficient of vapor ethanol ($D_v$) & $11.81 \times 10^{-6}$ m$ ^2 $/s \cite{lugg1968diffusion}\\
				Partial pressure of air ($P_a$) & $ 10^5$ Pa  \cite{mokeba1997simulating}\\
				Vapor pressure difference ($\Delta P$) & $ 790$ Pa  \cite{mokeba1997simulating}\\
				Kinematic viscosity of air ($\nu_a$) & $ 1.516 \times 10^{-5}$ m$ ^2 $/s \cite{mokeba1997simulating} \\
				\hline   \hline           
		\end{tabular}}
		\label{Sim_parameters}
	\end{table}
	\subsection{Results and Analysis}
	\label{Numerical_3}
	In Table \ref{Sim_parameters}, $Re$ is given as a very small number ($0.1$), since the medium between the TX and RX is assumed as non-turbulent. Furthermore, $\alpha_{d_0}$ is chosen as a small number ($0.001$) in accordance with the observations about the mixture volume of droplets and air particles.  Furthermore, the chemical properties of ethanol and air  in Table \ref{Sim_parameters} are obtained from \cite{pubchem_ethanol,lugg1968diffusion} and \cite{mokeba1997simulating}, respectively. $\Delta P$ is used as the same value in \cite{mokeba1997simulating}, since the densities of liquid in \cite{mokeba1997simulating} and our study have close values.

\begin{figure}[tbp]
		\centering
		\scalebox{0.45}{\includegraphics{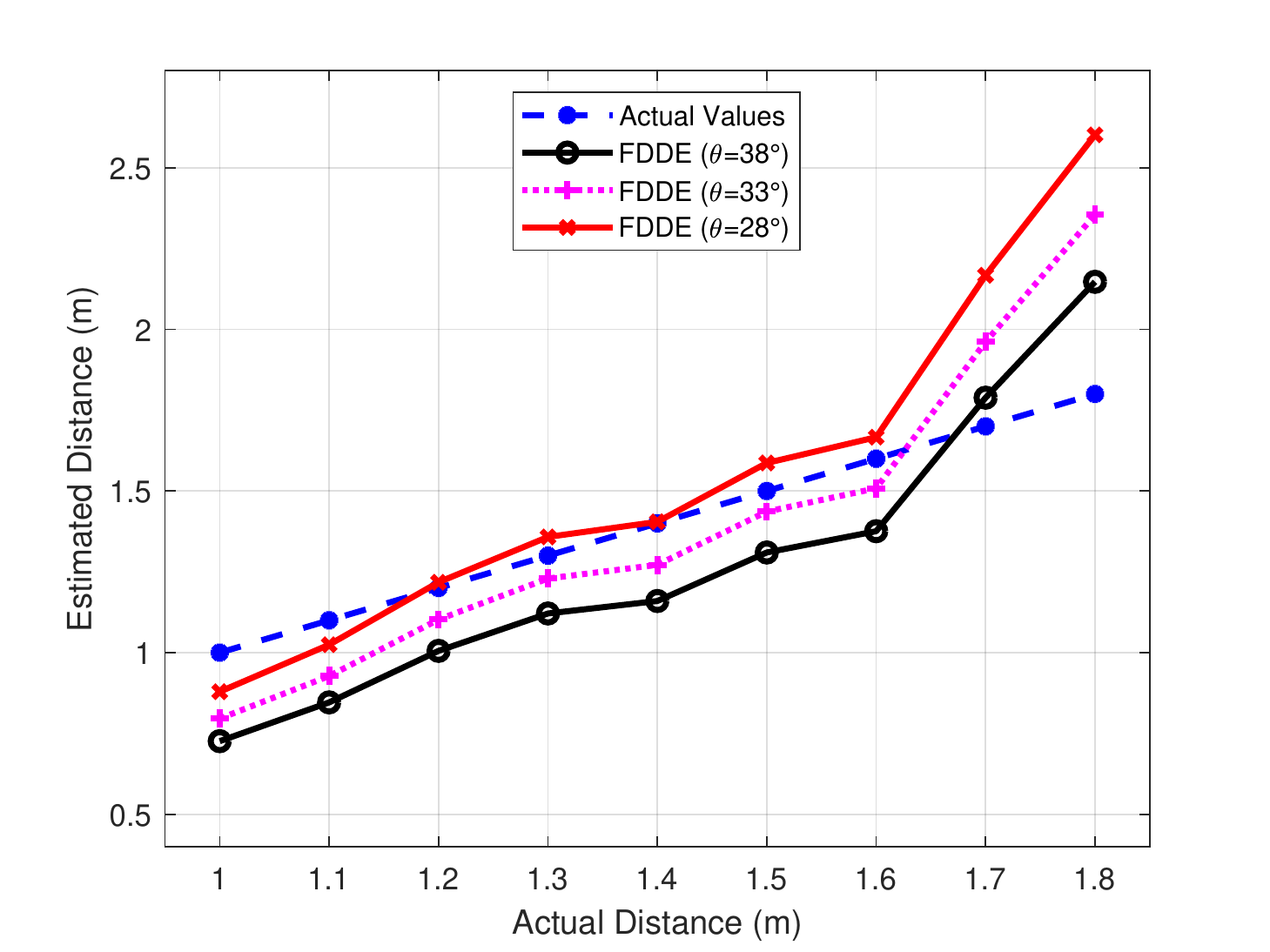}}  
		\caption{Experimental and estimated distance values with the FDDE algorithm.}
		\label{Est_s_plot}
	\end{figure}

\begin{figure}[b]
	\centering
	\scalebox{0.45}{\includegraphics{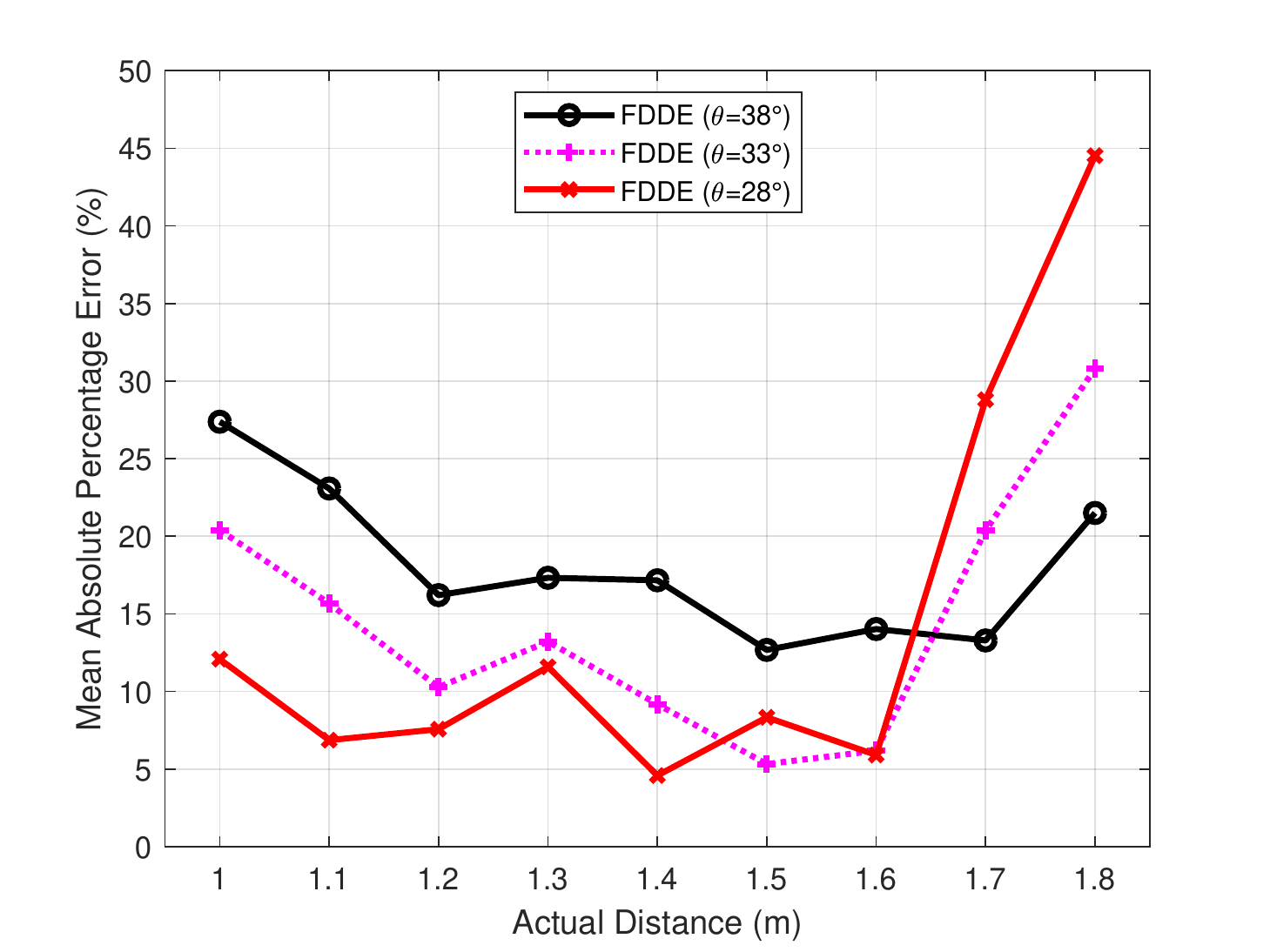}}  
	\caption{Mean Absolute Percentage Error between experimental and estimated distance values.}
	\label{MAPE_plot}
\end{figure}

	When the peak time of the sensor voltage, which is assumed as $t_i$, is measured by the RX, the distance can be estimated by the FDDE algorithm as shown in Fig. \ref{Est_s_plot}. Here, $\Delta t$ is chosen sufficiently small so that peak time values measured by the sensor correspond to $t_i$ values exactly. Each value in Fig. \ref{Est_s_plot} denotes the mean value of five estimations for the corresponding distance. In order to evaluate the accuracy of the estimations for each distance, Mean Absolute Percentage Error ($ \epsilon $) is chosen as the performance metric as given by
	\begin{equation}
	\epsilon = \frac{100}{N} \sum_{k=1}^{N} \frac{|\hat{s}_k - s|}{s}
	\end{equation}
	where $N$ is the number of estimations, $\hat{s}$ and $s$ are the estimated and actual distance values. Fig. \ref{MAPE_plot} shows the performance of the FDDE algorithm in terms of $\epsilon$. Figs. \ref{Est_s_plot} and \ref{MAPE_plot} show the effect of $\theta$ by using its measured value ($38^\circ$) and two different values ($ 33^\circ$, $28^\circ$). This effect is related with the physical phenomena about the spraying pattern and interaction of droplets with air particles. The sprayer affects the spraying pattern depending on the parameters such as $\theta$, the spatial dispersion and size of the droplets \cite{al2014influence}. Our observations during the experiments show that the majority of droplets propagate in a narrower beamwidth with respect to the measured value of $\theta$. Furthermore, the interactions among droplets and air particles induce an air flow in the vicinity of the nozzle towards the center of the beamwidth along the horizontal axis. This flow entrains droplets from the outer region into the inner region within the beamwidth \cite{ghosh1994induced}. This fact also supports the narrowing beamwidth after spraying droplets. Hence, as $\theta$ decreases, the FDDE algorithm gives more accurate results for shorter distances. The narrowing beamwidth can be estimated via high-speed photography techniques by visual analysis of the spray pattern or Phase Doppler Anemometry \cite{begg2009vortex}. However, the estimation and measurement of this narrowing beamwidth is beyond the scope of this paper.
	
	In Figs. \ref{Est_s_plot} and Fig. \ref{MAPE_plot}, the performance of the FDDE algorithm for each $\theta$ value changes dramatically for larger distances. This change can be explained with the effect of Brownian motion for the droplets at longer distances as also discussed in \cite{gulec2019distance}. The effect of Brownian motion is negligible for large droplets at smaller distances. However, as the distance increases, the average velocity of droplets is affected by Brownian motion due to decreasing droplet sizes. Therefore, the FDDE algorithm gives better results for shorter distances.

\begin{figure}[htbp]
		\centering
		\scalebox{0.45}{\includegraphics{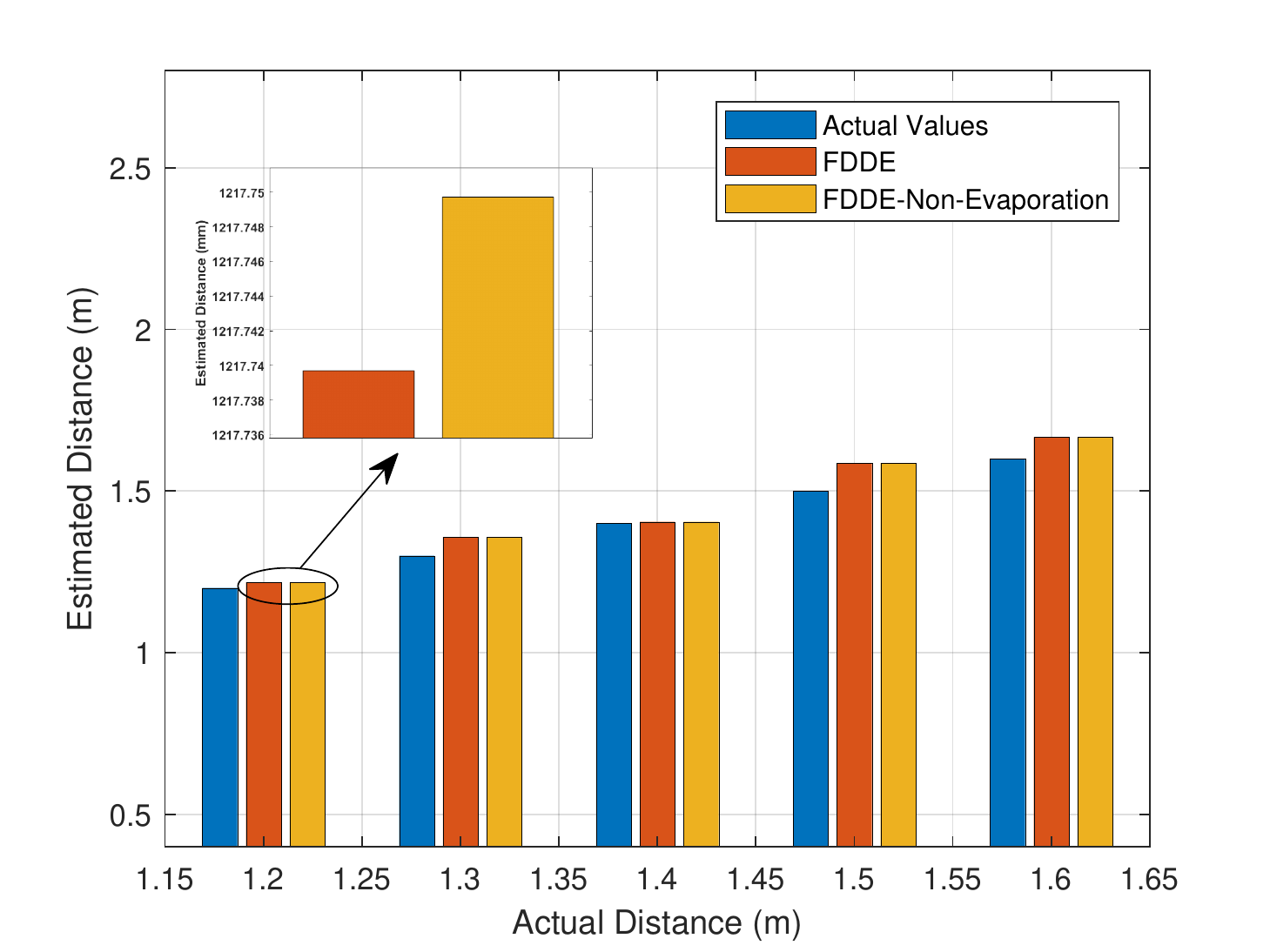}}  
		\caption{The effect of evaporation in FDDE algorithm.}
		\label{Evap_plot}
	\end{figure}

	Fig. \ref{Evap_plot} shows the effect of the droplet evaporation for distance estimation with the FDDE algorithm. The results in this figure for the non-evaporation case is obtained by ignoring the evaporation of droplets after $T_e$. For our experimental scenario, the effect of evaporation is small as it can be seen in magnified view of the results for $1.2$ m in Fig. \ref{Evap_plot}. The effect of evaporation for other distances in this figure are similar to the results obtained for $1.2$ m. Although this effect is very small for short distances, it can be more influential for longer distances and higher air temperatures due to the increment of the evaporation.
	\section{Conclusion}
	\label{Conclusion}
	In this paper, the FDDE algorithm which uses a fluid dynamics perspective is proposed for distance estimation in practical macroscale MC scenarios. For this algorithm, the propagation of transmitted droplets is modeled by using fluid dynamics  where it is assumed that droplets and air particles move together as a mixture, i.e., two-phase flow. Here, emitted droplets are modeled as evaporating spherical particles in the MC channel. The traveling distance of this mixture derived using this model is employed in the proposed algorithm to estimate the distance between the TX and RX. Afterwards, the FDDE algorithm is validated by experimental data. It is revealed that the distance depends on the initial velocity of droplets, air-droplet interaction in the two-phase flow, nozzle orifice, beamwidth of the TX and densities of droplet and air. In addition, the effect of droplet evaporation is shown with the numerical results. Hence, it is concluded that fluid dynamics approach can be employed to model the movements of droplets and estimate the parameters of a macroscale MC channel. As the future work, we plan to employ fluid dynamics approach for localization in a practical macroscale molecular network.
	
	%However, the model can be improved by taking the propagation of the evaporated particles into account. Furthermore, the evaporation model gives us information about the received molecular concentration at the RX due to the phase change of the droplets from liquid to gas in the molecular communication channel. This situation is needed to be considered for macroscale practical scenarios in molecular communication to clarify the propagation time and the signal shape of the received molecular concentration at the RX.

	% or
	

	%\appendices
	%% you can choose not to have a title for an appendix
	%% if you want by leaving the argument blank
	%\section{}
	%\allowdisplaybreaks
	%\appendix
	
	%\section*{Acknowledgment}
	%This work was supported by the Scientific and Technological Research Council of Turkey (TUBITAK) under Grant 119E041.
	
	%\ifCLASSOPTIONcaptionsoff
	%  \newpage
	%\fi
	
	\bibliographystyle{ieeetran}
	\bibliography{ref_fg_FDDE}

\begin{thebibliography}{10}
\providecommand{\url}[1]{#1}
\csname url@samestyle\endcsname
\providecommand{\newblock}{\relax}
\providecommand{\bibinfo}[2]{#2}
\providecommand{\BIBentrySTDinterwordspacing}{\spaceskip=0pt\relax}
\providecommand{\BIBentryALTinterwordstretchfactor}{4}
\providecommand{\BIBentryALTinterwordspacing}{\spaceskip=\fontdimen2\font plus
\BIBentryALTinterwordstretchfactor\fontdimen3\font minus
  \fontdimen4\font\relax}
\providecommand{\BIBforeignlanguage}[2]{{%
\expandafter\ifx\csname l@#1\endcsname\relax
\typeout{** WARNING: IEEEtran.bst: No hyphenation pattern has been}%
\typeout{** loaded for the language `#1'. Using the pattern for}%
\typeout{** the default language instead.}%
\else
\language=\csname l@#1\endcsname
\fi
#2}}
\providecommand{\BIBdecl}{\relax}
\BIBdecl

\bibitem{nakano2013molecular}
T.~Nakano, A.~W. Eckford, and T.~Haraguchi, \emph{Molecular
  communication}.\hskip 1em plus 0.5em minus 0.4em\relax Cambridge University
  Press, 2013.

\bibitem{Atakan-2014}
B.~Atakan, \emph{Molecular communications and nanonetworks: from nature to
  practical systems}.\hskip 1em plus 0.5em minus 0.4em\relax Springer Science
  \& Business Media, 2014.

\bibitem{Farsad-2016}
N.~Farsad, H.~B. Yilmaz, A.~Eckford, C.-B. Chae, and W.~Guo, ``A comprehensive
  survey of recent advancements in molecular communication,'' \emph{IEEE
  Commun. Surv. \& Tut.}, vol.~18, no.~3, pp. 1887--1919, 2016.

\bibitem{farsad2013tabletop}
N.~Farsad, W.~Guo, and A.~W. Eckford, ``Tabletop molecular communication: Text
  messages through chemical signals,'' \emph{PloS one}, vol.~8, no.~12, p.
  e82935, 2013.

\bibitem{farsad2017novel}
N.~Farsad, D.~Pan, and A.~Goldsmith, ``A novel experimental platform for
  in-vessel multi-chemical molecular communications,'' in \emph{IEEE GLOBECOM},
  2017, pp. 1--6.

\bibitem{unterweger2018experimental}
H.~Unterweger~et.al., ``Experimental molecular communication testbed based on
  magnetic nanoparticles in duct flow,'' in \emph{2018 IEEE 19th SPAWC}, pp.
  1--5.

\bibitem{giannoukos2017molecular}
S.~Giannoukos, A.~Marshall, S.~Taylor, and J.~Smith, ``Molecular communication
  over gas stream channels using portable mass spectrometry,'' \emph{J. Amer.
  Soc. Mass Spectrometry}, vol.~28, no.~11, pp. 2371--2383, 2017.

\bibitem{mcguiness2018parameter}
D.~T. McGuiness, S.~Giannoukos, A.~Marshall, and S.~Taylor, ``Parameter
  analysis in macro-scale molecular communications using advection-diffusion,''
  \emph{IEEE Access}, vol.~6, pp. 46\,706--46\,717, 2018.

\bibitem{koo2016molecular}
B.-H. Koo, C.~Lee, H.~B. Yilmaz, N.~Farsad, A.~Eckford, and C.-B. Chae,
  ``Molecular mimo: From theory to prototype,'' \emph{IEEE J. on Sel. Areas in
  Commun.}, vol.~34, no.~3, pp. 600--614, 2016.

\bibitem{zhai2018anti}
H.~Zhai, Q.~Liu, A.~V. Vasilakos, and K.~Yang, ``Anti-isi demodulation scheme
  and its experiment-based evaluation for diffusion-based molecular
  communication,'' \emph{IEEE Trans. on Nanobioscience}, vol.~17, no.~2, pp.
  126--133, 2018.

\bibitem{zhai2018bio}
H.~Zhai, L.~Yang, T.~Nakano, Q.~Liu, and K.~Yang, ``Bio-inspired design and
  implementation of mobile molecular communication systems at the macroscale,''
  in \emph{IEEE GLOBECOM}, 2018, pp. 1--6.

\bibitem{atakan2007information}
B.~Atakan and O.~B. Akan, ``An information theoretical approach for molecular
  communication,'' in \emph{BIONETICS}, 2007, pp. 33--40.

\bibitem{nakano2013transmission}
T.~Nakano, Y.~Okaie, and A.~V. Vasilakos, ``Transmission rate control for
  molecular communication among biological nanomachines,'' \emph{IEEE J. Sel.
  Areas in Commun.}, vol.~31, no.~12, pp. 835--846, 2013.

\bibitem{khalid2019communication}
M.~Khalid, O.~Amin, S.~Ahmed, B.~Shihada, and M.-S. Alouini, ``Communication
  through breath: Aerosol transmission,'' \emph{IEEE Commun. Mag.}, vol.~57,
  no.~2, pp. 33--39, 2019.

\bibitem{moore2010measuring}
M.~Moore, T.~Nakano, A.~Enomoto, and T.~Suda, ``Measuring distance with
  molecular communication feedback protocols,'' \emph{Proc. ICST BIONETICS},
  pp. 1--13, 2010.

\bibitem{moore2012measuring}
M.~J. Moore, T.~Nakano, A.~Enomoto, and T.~Suda, ``Measuring distance from
  single spike feedback signals in molecular communication,'' \emph{IEEE Trans.
  on Signal Process.}, vol.~60, no.~7, pp. 3576--3587, 2012.

\bibitem{moore2012comparing}
M.~J. Moore and T.~Nakano, ``Comparing transmission, propagation, and receiving
  options for nanomachines to measure distance by molecular communication,'' in
  \emph{IEEE ICC}, 2012, pp. 6132--6136.

\bibitem{huang2013distance}
J.-T. Huang, H.-Y. Lai, Y.-C. Lee, C.-H. Lee, and P.-C. Yeh, ``Distance
  estimation in concentration-based molecular communications,'' in \emph{IEEE
  GLOBECOM}, 2013, pp. 2587--2591.

\bibitem{wang2015distance}
X.~Wang, M.~D. Higgins, and M.~S. Leeson, ``Distance estimation schemes for
  diffusion based molecular communication systems,'' \emph{IEEE Commun. Lett.},
  vol.~19, no.~3, pp. 399--402, 2015.

\bibitem{wang2015algorithmic}
------, ``An algorithmic distance estimation scheme for diffusion based
  molecular communication systems,'' in \emph{IEEE ICC}, 2015, pp. 1134--1139.

\bibitem{lin2019high}
L.~Lin, Z.~Luo, L.~Huang, C.~Luo, Q.~Wu, and H.~Yan, ``High-accuracy distance
  estimation for molecular communication systems via diffusion,'' \emph{Nano
  Commun. Netw.}, vol.~19, pp. 47--53, 2019.

\bibitem{noel2015joint}
A.~Noel, K.~C. Cheung, and R.~Schober, ``Joint channel parameter estimation via
  diffusive molecular communication,'' \emph{IEEE Trans. on Mol., Biol. and
  Multi-Scale Commun.}, vol.~1, no.~1, pp. 4--17, 2015.

\bibitem{gulec2019distance}
F.~Gulec and B.~Atakan, ``Distance estimation methods for a practical
  macroscale molecular communication system,'' \emph{arXiv:1909.12897}, 2019.

\bibitem{ghosh1994induced}
S.~Ghosh and J.~C.~R. Hunt, ``Induced air velocity within droplet driven
  sprays,'' \emph{Proc. R. Soc. Lond. A}, vol. 444, no. 1920, pp. 105--127,
  1994.

\bibitem{al2014influence}
M.~Al~Heidary, J.~Douzals, C.~Sinfort, and A.~Vallet, ``Influence of spray
  characteristics on potential spray drift of field crop sprayers: A literature
  review,'' \emph{Crop protection}, vol.~63, pp. 120--130, 2014.

\bibitem{sazhin2001model}
S.~Sazhin, G.~Feng, and M.~Heikal, ``A model for fuel spray penetration,''
  \emph{Fuel}, vol.~80, no.~15, pp. 2171--2180, 2001.

\bibitem{mokeba1997simulating}
M.~Mokeba, D.~Salt, B.~Lee, and M.~Ford, ``Simulating the dynamics of spray
  droplets in the atmosphere using ballistic and random-walk models combined,''
  \emph{J. Wind Eng. and Ind. Aerodynamics}, vol.~67, pp. 923--933, 1997.

\bibitem{munson2009fundamentals}
B.~R. Munson, D.~F. Young, T.~H. Okiishi, and W.~W. Huebsch, \emph{Fundamentals
  of fluid mechanics}.\hskip 1em plus 0.5em minus 0.4em\relax John Wiley \&
  Sons, Inc, 2009.

\bibitem{pubchem_ethanol}
``Chemical properties of ethanol,'' \url{https://pubchem.ncbi.nlm.nih.gov},
  accessed: 2019-10-01.

\bibitem{lugg1968diffusion}
G.~Lugg, ``Diffusion coefficients of some organic and other vapors in air,''
  \emph{Analytical Chemistry}, vol.~40, no.~7, pp. 1072--1077, 1968.

\bibitem{begg2009vortex}
S.~Begg, F.~Kaplanski, S.~Sazhin, M.~Hindle, and M.~Heikal, ``Vortex ring-like
  structures in gasoline fuel sprays under cold-start conditions,'' \emph{Int.
  J. Engine Res.}, vol.~10, no.~4, pp. 195--214, 2009.

\end{thebibliography}
	
\end{document}